# Optimizing Distributed Networking with Big Data Scheduling and Cloud Computing

Wenbo Zhu
College of Arts and Science
New York University
New York, NY10003, USA
wz1305@nyu.edu

*Abstract*—With the rapid transformation of computer hardware and algorithms, mobile networking has evolved from low data-carrying capacity and high latency to better-optimized networks, either by enhancing the digital network or using different approaches to reduce network traffic. This paper discusses the big data applications and scheduling in the distributed networking and analyzes the opportunities and challenges of data management systems. The analysis shows that the big data scheduling in the cloud computing environment produces the most efficient way to transfer and synchronize data. Since scheduling problems and cloud models are very complex to analyze in different settings, we set it to the typical software defined networks. The development of cloud management models and coflow scheduling algorithm is proved to be the priority of the digital communications and networks development in the future.

## I. INTRODUCTION

Network is everywhere in the world, and it exists not only in the communities that people are living in, but also exists at the space station that beyond our reach. The networking infrastructures has changed through the past decades, more and more servers and stations are built to extend and amplify the signal of different networks. Networks such as cellular networks or WIFI has now become the most essential communications way that people get in touch with each other in their daily life. Since the networks interfere with our social interactions, the low latency and high speed of the networks are required for the communications. The mobility and scalability of the network act as a milestone for people to conquer, and research efforts has been pushing the efficiency of the transmission of data to a higher level. The forthcoming big data based digital networks have gained a lot of attention from various industries, and pioneering theoretical and practical improvement can have an unprecedented effect on current networking infrastructures. The usage of cloud computing, similar to a management system, is a hopeful field that allows users to have more storage and processing capacity, which can accelerate or even revolutionize the development of mobile networking services. The distributed networking, as a central networking solution for the more advanced broadband communication and diverse applications in the next decades, played an important role to optimize the current complex network structure and revolutionize the methodology of transferring data. The big data scheduling and cloud computing environment have huge effect on the technological innovation in distributed networking.

## II. BIG DATA ANALYTICS

### A. Brief Introduction

The amount of data produced by people's daily life in terms of user interfaces and online social interactions has more than doubled in recent years. These large volumes of data have been identified and categorized as "Big Data". Hace Theorem stated that big data is heterogeneous and diverse with distributed control, and it further illustrates a vivid picture that big data can be considered as a giant elephant that combines heterogeneous and autonomous data [1]. The complex and constantly changing nature of data made it not possible to grasp the whole picture of the field of big data.

### B. Advantages

Applications of big data has been approved in industrial and commercial field, and it was advanced by different researchers in order to process various information in an efficient manner. While researchers also explore the theoretical possibilities of the real-world usage of big data, it has been mainly motivated by the commercial or political use of data mining and machine learning [2]. Big data is better for the reconstruction and optimization of the data mining process and machine learning models. The data management system that optimized by big data analytics, is allowed to have higher uploading speed and lower latency in the networks. In addition, the network traffic can be avoided in terms of scheduling data through different internet connections in the heterogeneous network architecture using data mining and machine learning techniques.

### C. Data Mining and Machine learning

Data minning and Machine learning models often require homogenous and centralized data, but big data deals with heterogeneous and decentralized data. Seeking common ground is the best way for the application of big data, and big data mining can be beneficial for building the framework of machine learning algorithms.

## III. DISTRIBUTED NETWORKING IN BIG DATA

Nowadays, data are distributed over different data centers just as the Internet is provided through various networking centers. Distributed networking for big data has emerged as a hot trend that there are newly established data and cloud computing centers for processing different types of data in correspondence to the particular software-defined networks such as OpenFlow [3].

*A. Networking for Big Data Scheduling*

In the networking for big data, while processing each level of the tasks, the way how we manage to schedule different layers of the tasks is important to the performance of the application. Distributed flow within a task is the key to big data scheduling and the traditional scheduling algorithms are working at a per-flow level [4]. Therefore, researchers proposed Inter-coflow scheduling to improve application performance using prior coflow information that traverses the different networks at different times. In the table below, it states the different method of coflow scheduling, which can be used to determine the prior knowledge of Inter-coflow [5].

TABLE I. AVERAGE SCHEDULING TIME USING DIFFERENT APPROACHES

| Approaches | Coflow-aware | Performance metrics | Routing | Priority knowledge of coflow |
|---|---|---|---|---|
| Pfabric [130] | No | Single flow completion time | No | – |
| Orchestra [131] | Yes | Coflow completion time | No | Yes |
| Baraat [28] | Yes | Average/Tail coflow completion time | No | No |
| Varys [29] | Yes | Decrease communication time of data-intensive jobs guarantee predictable communication time | No | Yes |
| D-CAS [132] | Yes | Average coflow completion time | No | Yes |
| Rapier [133] | Yes | Routing Average coflow completion time | Yes | Yes |
| Aalo [134] | Yes | Average coflow completion time | No | No |

*B. Cybersecurity in networking for Big Data—Software-Defined Networking*

It is well known that people are concerning the violation of privacy and security when signing a term of agreement online, and the online application is drawing the attention of big data security with networking.

In distributed networking systems, one specific type of network is reviewed and analyzed. The SDN (Software-Defined Networking) is able to provide services to applications that are based on the diverse nature of big data. Such SDN protocol as OpenFlow would have the ability to centralize resources and distribute information efficiently. The SDN is innovative that allows heterogeneous data representation in a large form.

## IV. CLOUD COMPUTING ON DISTRIBUTED NETWORKING

*A. Brief Introduction*

In the 21st century, computing power has been increasing dramatically, and cloud computing is the suitable way that the computing power is mainly supplied. As people are desperate in need of lightweight portable devices that function from the user end, it requires more and more computing powers. The ubiquitous wireless networking demand low storage costs with high performance.

*B. Advantages*

The cloud computing infrastructure typically located in the large management center which considers the power efficiency of services. On-demand computing supports the combination of software, hardware, and data to deliver services to the user.

*1) Data Management Systems*

The multi-tenant database design deploys multiple data management tasks to reduce delay and improve cloud computing capabilities. In cloud computing infrastructures, there are still many ongoing challenges and opportunities to optimize the systems such as the Data Management systems. Data management deals with the scalability of databases which controls the changing demand by inserting or deleting resources dynamically [6]. Cloud computing would give distributed databases more computing power in remote locations, which records the availability of the data management systems and relocates resources based on the performance of the system.

*2) Parallel programming model—MapReduce-like systems*

In the data centers, MapReduce-like systems run through different original data sets, and output two functions consist of Map and Reduce that can happen in the synchronized time. Optimizing MapReduce-like systems in the cloud computing environment would reform the assignment and scheduling function that minimizes the completion time of all jobs in the cluster. Different scheduling strategies are listed below in the picture, which performs each simulation with three different scheduling algorithms called SJF, STF, FIFO in the simulation setting [7].

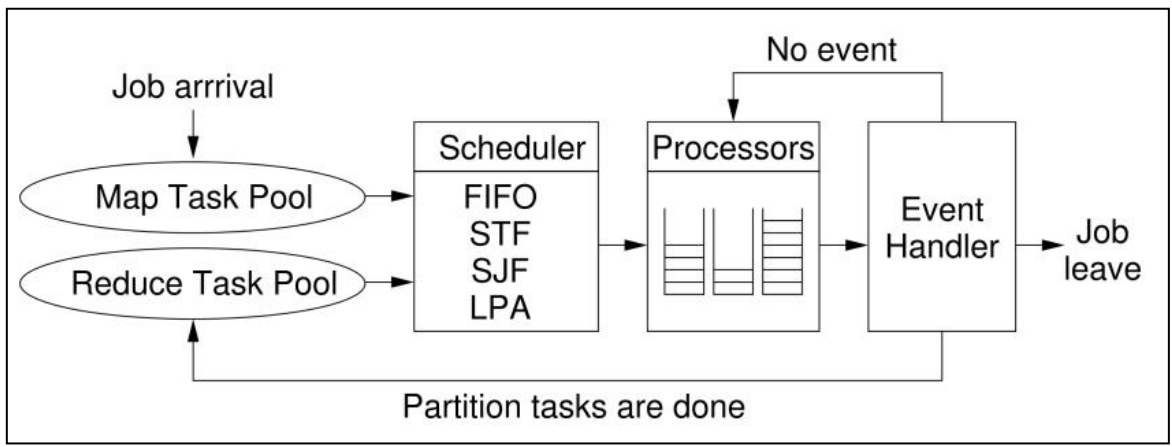

Figure 1.  The simulation of MapReduce System

*3) Privacy and Security of SDN in the cloud computing environment*

In the typical distributed networking environment, SDN is one of the networks that we have commonly seen in our daily life. The privacy and security of SDN are vital to our day-to-day life, and the characteristics of SDN made it easy to detect and react to internet attacks such as denial-of-service (DDoS) attacks [8]. There is an increasing trend of newly DDoS attacks, and make use of SND's characteristics is the best way to protect the user from the attacks. The on-demand services, broad networking pooling, and networking elasticity are the main target of the DDoS attacks in the cloud computing environment, the resource pooling in the SDN could use its own features to defeat the DDoS attacks shown in the picture below [8].

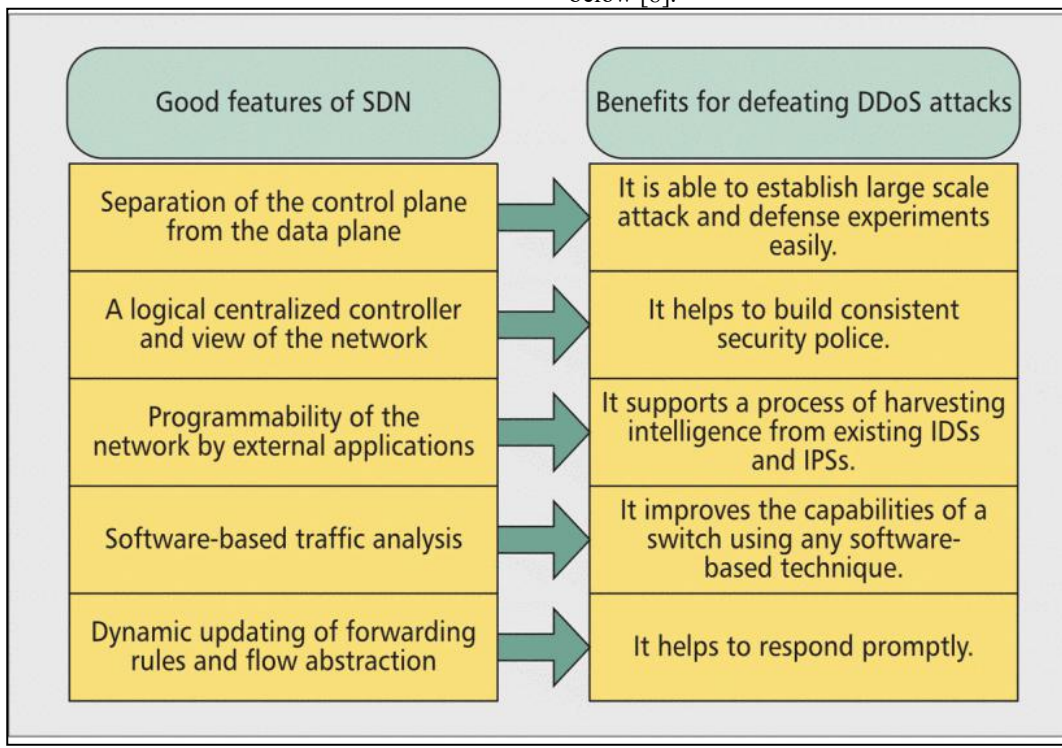

Figure 2.  SDN on DDoS attack

*4) Distributed Storage*

Google File System(GFS) and Hadoop Distributed File System(HDFS) are two data storage systems that are mainly used for cloud computing. Mass distributed storage and reliable computing system are invented and improved through the race of developing cloud computing systems. While GFS runs on cheap and common hardware, HDFS runs on commodity hardware which is similar to the existed distributed file system. GFS exchanges information between the master server and block server in the cloud and HDFS used the architecture of Master/Slave that makes up Namenode and Datanode in which they operate the namespace and dataspace in the mapping system [9].

## V. Discussion on Future trends of distributed networking

### A. Big Data Analytics

Many online applications in software-defined networking are data-oriented, and they rely on the wireless communication between different sensor nodes in the distributed system. The fundamental networking technologies require a lot of data that gathers from the usage of Big Data. The analytics of big data would accelerate the process and improve data locality on a large scale.

For the purpose of synchronizing data with different collaborators, a lot of data management centers have emerged to achieve scalable in-place data sharing such as Globus. Sharing data with collaborators in synchronize time on the cloud uses Grid Security Infrastructure(GSI) to allow fast and secure communications [10]. Cloud computing would process data in the endpoint, while GSI establishes secure connections and provide trusted certificates for users to access the server.

### B. Cloud Computing Systems

Data management and MapReduce system have great effects on the cloud computing environment, thus the development of that system should be promoted as the priority of the field of data science. Distributed storage systems are the solutions for the application of massive datasets in our daily life, the combination of data management systems and file management systems would enable users to use cloud computing resources efficiently. Cloud computing infrastructures still have a lot of room for improvement, for instance, the privacy and security of the user and power efficiency of the data management centers, which requires constant and consistent efforts from academic and industrial researchers. Optimizing the data management algorithm, implementing models in distributed storage, and synchronizing in-Place data sharing in the cloud computing environment is the essential way to make real-world applications in practice.

### C. Efficient Synthesis of Networking Environment

As more and more data flows in and out of the internet, the usage of big data and cloud computing in distributed networking comes to an optimal place to develop in the private and public sectors. The local and remote pool of resources in the internal and external clouds act as a focusing point for companies to deliver online services in the distributed storage-based data management systems. In the stage of developing virtualized distributed computing infrastructure, and the scalability of the infrastructure would show the world about the benefit of running applications on the cloud-based platform. Monetizing cloud computing systems and big data applications are probably favorable by the companies that build internet consumer services, and pushing the field to a higher level.

## VI. Conclusion

Distributed networking in big data allows people to have various networking connections through global networking centers. Particularly, the software-defined networks allow people to distinguish diverse data in terms of representing them in a large unity, and big data coflow scheduling can increase the performance of the application.

The data-oriented SDN is suitable to process data in the big data background. Data management systems such as MapReduce system in the cloud computing environment provide synchronizing functions in the cluster of data. The distributed storage of data in the file management system would protect users' privacy and security in an efficient way. Synchronizing big data applications with cloud computing capacity in the distributed form would provide the user a better experience with these data management infrastructures by the embedded advanced data management algorithm and innovative models in the cloud-based platform. However, the experimental big data scheduling on different cloud models with the software defined networks was not hypothesized and tested in the paper, and the analysis related to those distributed networks model might be studied in the following research.

Acknowledgment

I would like to express my gratitude to every single author of the reference, this work is impossible to complete without your previous efforts. This article was completed under the guidance of Professor Jean Franchitti, Instructor Jane, Han, and Katherine. I appreciate the guidance in distributing networks from Professor Franchitti and Instructor Jane, which broad my selection of content and writing specifications. At the same time, I thank Instructor Katherine for supervising me and providing instructions on the paper. At last, I would like to thank again everyone who has helped me with the paper, and your assistance has made this work possible and valuable.

References

[1] X. Wu, X. Zhu, G. Wu and W. Ding, "Data mining with big data," in IEEE Transactions on Knowledge and Data Engineering, vol. 26, no. 1, pp. 97-107, Jan, 2014, doi: 10.1109/TKDE.2013.109.

[2] M. Jordan and T. MItchell, "Machine learning: trends, perspectivs, and prospects," Science, vol. 349, no. 6245, pp. 255–260, 2015.

[3] N. McKeown, T. Anderson, H. Balakrishnan, G. Parulkar, L. Peterson, J. Rexford, S. Shenker, and J. Turner, "Openflow: enabling innovation in campus networks," ACM SIGCOMM Computer Communication Review, vol. 38, no. 2, pp. 69–74, 2008.

[4] C. Y. Hong, M. Caesar, and P. Godfrey, "Finishing flows quickly with preemptive scheduling," ACM SIGCOMM Computer Communication Review, vol. 42, no. 4, pp. 127–138, 2012.

[5] Yu, S., Liu, M., Dou, W., Liu, X., & Zhou, S. Networking for Big Data: A Survey. IEEE Communications Surveys & Tutorials, 19(1), pp. 531–549, 2017, doi:10.1109/comst.2016.2610963

[6] Agrawal, D., El Abbadi, A., Antony, S., & Das, S. Data Management Challenges in Cloud Computing Infrastructures. Lecture Notes in Computer Science, 2010, pp. 1–10. doi:10.1007/978-3-642-12038-1_1

[7] Chang, H., Kodialam, M., Kompella, R. R., Lakshman, T. V., Lee, M., & Mukherjee, S. Scheduling in mapreduce-like systems for fast completion time. 2011 Proceedings IEEE INFOCOM. 2011, doi:10.1109/infcom.2011.5935152

[8] Yan, Q., & Yu, F. R. Distributed denial of service attacks in software-defined networking with cloud computing. IEEE Communications Magazine, 53(4), pp. 52–59, 2015, doi:10.1109/mcom.2015.7081075

[9] Zhang, S., Yan, H., & Chen, X. Research on Key Technologies of Cloud Computing. Physics Procedia, 33, pp. 1791–1797, 2012, doi:10.1016/j.phpro.2012.05.286

[10] Chard, K., Tuecke, S., & Foster, I. Efficient and Secure Transfer, Synchronization, and Sharing of Big Data. IEEE Cloud Computing, 1(3), pp. 46–55, 2014, doi:10.1109/mcc.2014.52